\documentclass[aps,pra,reprint,superscriptaddress,showpacs]{revtex4-1}

\usepackage{amsmath,amssymb,amstext}
\usepackage{mathrsfs}
\usepackage{dsfont}
\usepackage{graphicx}
\usepackage[all]{xy}

\newcommand{\NN}{{\mathbb N}}

\newcommand{\eps}{\varepsilon}

\newcommand{\ket}[1]{\left \vert #1 \right \rangle}
\newcommand{\braket}[2]{\langle #1 \vert #2 \rangle}
\newcommand{\matel}[3]{{\left\langle \vphantom{#1 #2 #3} #1 \,\right\vert
\left.
 \hspace{-0.15em} \vphantom{#1 #2 #3} #2 \,\right\vert \left.
 \hspace{-0.15em} \vphantom{#1 #2 #3} #3\right\rangle}}

\newcommand{\ketbra}[2]{\vert #1 \rangle \langle #2 \vert}

\newcommand{\be}{\begin{equation}}
\newcommand{\ee}{\end{equation}}
\newcommand{\bea}{\begin{eqnarray}}
\newcommand{\eea}{\end{eqnarray}}

\begin{document}

\title{Molecular Rotations in Matter-Wave Interferometry}

\author{Benjamin A. Stickler}
\email{benjamin.stickler@uni-due.de}
\author{Klaus Hornberger}

\affiliation{Faculty of Physics, University of Duisburg - Essen, Lotharstra\ss e 1, 47048 Duisburg}


\begin{abstract}
We extend the theory of matter-wave interferometry of point-like particles to non-spherical objects by taking the orientational degrees of freedom into account. In particular, we derive the grating transformation operator, that maps the impinging state onto the outgoing state, for a general, orientation-dependent interaction potential between the grating and the molecule. The grating operator is then worked out for the diffraction of symmetric top molecules from standing light waves, and the resulting interference pattern is calculated in the near-field. This allows us to identify a signature of the orientational degrees of freedom in near-field matter-wave experiments.
\end{abstract}

\maketitle

\section{Introduction}

Interference experiments with heavy molecules and nanoclusters amount to tests of the quantum superposition principle at unprecedented scales \cite{haslinger2013universal,colloquium,romeroisart2011,stefanjames} and may serve to measure molecular properties with high accuracy \cite{conformationspra,sandraabsoprtion,hackermuller2007optical}. While it is an open question whether the superposition principle is valid at all scales \cite{bassi,macroscopicity,arndt2014testing}, it is without doubt that the ro-vibrational degrees of freedom of increasingly large, non-spherical objects will at some point influence the interference signal.

Far-field matter-wave experiments with heavy particles are challenging due to the small de Broglie wavelength, but near-field interferometry proved to be a powerful tool in the quest for high mass interference \cite{kdtlinature,kdtlinjp,haslinger2013universal,otimanjp,colloquium}. Near-field techniques are based on the Talbot effect, the reproduction of the intensity pattern in the grating at certain distances further downstream. Since near-field interference effects are highly sensitive to even tiny phase modifications, it is natural to expect that the influence of the rotational state is most pronounced in the near field. In fact, signatures of the vibrational molecular dynamics have been observed in near-field experiments in \cite{conformationspra}.

In order to extend the established theory of matter-wave interferometry of spherical particles to non-spherical objects with orientational degrees of freedom, we draw on the results obtained for the deflection of rotating molecules \cite{averbukh2010,averbukh2011a,averbukh2011b,seideman05,stapelfeldt03,denschlag14}. There, it is a central result that the deflection angle of a rapidly rotating molecule is determined by its rotational state when entering the deflection field. In most cases of interest, the rotation of the molecule is initially thermally distributed, in particular in the absence of a pre-aligning pulse. The distribution of the rotational state then translates directly into a range of deflection angles, each observed with the thermal probability of the respective state. The distribution describing the probability of a particular deflection angle is thus a central element in the theory of molecular deflection and it will turn out to be similarly relevant for the theory of 
matter-wave interferometry of rotating molecules.

This article presents a full quantum theory of the influence of rotations in matter-wave interferometry, and we illustrate this theory by calculating the near-field interference pattern of symmetric top molecules in the Kapitza-Dirac-Talbot-Lau interferometer (KDTLI). We emphasize that the general formalism presented here is not restricted to the orientation state, but can be applied to other internal degrees of freedom as well.

The article is structured as follows: In Sec. \ref{sec:grattrafo} we derive the quantum mechanical grating transformations for the  cases of a rotationally free and a rotationally diabatic transit through the grating as well as their classical analogues. In Sec. \ref{sec:example} we specify the grating transformation for the diffraction of symmetric top molecules from a standing light wave in order to study the near-field interference pattern in the Kapitza-Dirac-Talbot-Lau interferometer (KDTLI), and we identify a signature of the rotational dynamics. We conclude in Sec. \ref{sec:conc}. The appendix provides a derivation of the distribution of classical deflection angles of rotating symmetric molecules, as required in Sec. \ref{sec:example}.

\section{The Grating Transformation} \label{sec:grattrafo}

It is the aim of this section to determine the effect of the grating on the translational and rotational dynamics of the molecule. For this sake we consider a molecule of mass $M$ that traverses with constant velocity $v_z$ a diffraction grating located at $z = 0$ and with the grating axis along the $x$-direction. The assumption that $v_z$ remains constant throughout the diffraction process is well justified because in the experimental realization the kinetic energy of the longitudinal motion exceeds the average interaction potential and the kinetic energy of the transverse motion by orders of magnitude \cite{stefanbeyond}. In addition, the restriction to a single velocity $v_z$ is not a limitation because a finite longitudinal coherence can always be incorporated by averaging over the distribution of $v_z$ in the end \cite{kdtlinjp}.

The orientation-dependent interaction between the molecule and the grating is described by the grating potential $V(x,z,\Omega)$, where $\Omega$ denotes the set of orientational degrees of freedom (DOFs) of the molecule, such as the Euler angles, $z = v_z t$ is the center-of-mass (CM) position of the molecule in the flight direction at time $t$ and $x$ is the CM position in the grating direction. Since the extension of the particle beam in $y$-direction is usually small compared to the extension of the grating, we can neglect the $y$-dependence of the potential $V(x,v_zt,\Omega)$.

In what follows, we distinguish between the rotationally \emph{free} transit, where the (expectation value of the) rotational period $\tau_{\rm rot}$ of the molecule is constant during the passage and much smaller than the transit time $\tau_{\rm CM}$ through the grating, and the rotationally \emph{diabatic} scenario, in which the characteristic time $\tau_{\rm CM}$ is much smaller than the rotational period $\tau_{\rm rot}$. While the rotationally free transit is realized, for instance, in near-field matter-wave interferometry with laser gratings \cite{kdtlinjp}, the diabatic transit can occur in far-field experiments with very thin material gratings \cite{naturephthalo}.

\subsection{Quantum Mechanical Description} \label{sec:quth}

The central tool in the theory of matter-wave interferometry of spherically symmetric particles is the grating transformation operator $\hat t$, that maps the incoming transverse state $\rho$ onto the outgoing transverse state $\rho' =  \hat t\rho \hat t^\dagger$ with $\matel{x}{\rho'}{x'} = t(x) t^*(x') \matel{x}{\rho}{x'}$ \cite{kdtlinjp,otimanjp,stefanbeyond}. In the case of extended, non-spherical molecules, the grating transformation operator $\hat t$ must be adapted in order to account for the effect of the orientational DOFs. Let us start by deriving this operator from the time-dependent Schr\"{o}dinger equation.

The total Hamiltonian $\hat H$ contains, in addition to the grating potential $V(x,z,\Omega)$, the CM kinetic energy $\hat H_{\rm CM}$ in transverse direction $x$, as well as the free rotational Hamiltonian $\hat H_{\rm rot}$, whose form is determined by the symmetry of the molecule \cite{sakurai}. In what follows we keep the discussion general and denote by $\ket{\ell m}$ the eigenfunctions of the rotational Hamiltonian with energy $\eps_{\ell}$, where $\ell$ labels the energy levels and $m$ labels the degenerate states for each $\ell$. The symmetry of the rotor is arbitrary and so is the degeneracy $\nu_\ell$ for each $\ell$.

For the sake of a more compact notation we introduce for each $\ell$ the tuple ${\underline \phi}_\ell(\Omega)$ containing all the states with energy $\eps_\ell$, i.e. $({\underline \phi_{\ell}})_m = \braket{\Omega}{\ell m}$. The length of the tuple ${\underline \phi}_\ell(\Omega)$ is thus equal to $\nu_\ell$. With these tuples the total wave function $\Psi(x,\Omega, t) = \braket{x,\Omega}{\Psi(t)}$ can be expanded as
\be \label{eq:exp}
\Psi(x,\Omega,t) = \sum_\ell  e^{-i \eps_\ell t/ \hbar} {\underline \chi}_\ell(x,t)\cdot{\underline \phi}_\ell(\Omega).
\ee
where ${\underline \chi}_\ell(x,t)$ are the tuples of expansion coefficients $({\underline \chi}_\ell)_m = \chi_{\ell m}$. Again, the length of ${\underline \chi}_\ell(x,t)$ is $\nu_\ell$. Inserting the expansion \eqref{eq:exp} into the time dependent Schr\"{o}dinger equation with the total Hamiltonian $\hat H$ gives the coupled equations
\bea \label{eq:preeffschr}
i \hbar \partial_t {\underline \chi}_\ell(x,t) & = & \hat H_{\rm CM} {\underline \chi}_\ell(x,t) \notag \\
 && + \sum_{\ell'}e^{- i \Delta_{\ell' \ell}  t / \hbar} {\underline {\underline V}}_{\ell \ell'}\left ( x, v_z t \right ) {\underline \chi}_{\ell'}(x,t).
\eea
Here, $\Delta_{\ell'\ell} = \eps_{\ell'} - \eps_\ell$ is the rotational energy-level spacing and we defined the grating potential matrix $({\underline {\underline V}}_{\ell \ell'})_{mm'}(x,v_z t) = \matel{\ell m}{V(x,v_z t,\hat \Omega)}{\ell' m'}$ with dimension $\nu_\ell \times \nu_{\ell '}$. In addition, we denote the initial conditions to the Schr\"{o}dinger equation \eqref{eq:preeffschr} by variables without the time argument, such as ${\underline \chi}_{\ell}(x)$.

Equation \eqref{eq:preeffschr} describes the coupled time evolution of the expansion coefficients $\chi_{\ell m}(x,t)$ due to the effectively time-dependent interaction between the molecule and the grating. The exact grating transformation $\hat t$ is given by the unitary time evolution of the system \eqref{eq:preeffschr}; however, in practice a semiclassical approach is sufficient due to the small de Broglie wavelength in matter-wave experiments with heavy molecules \cite{stefanbeyond}. In most cases it is even sufficient to determine the grating transformation $\hat t$ in the eikonal approximation, which can be regarded as the high energy limit of the semiclassical propagator \cite{stefanbeyond}. Physically speaking, the eikonal approximation treats the particle trajectories appearing in the semiclassical propagator as straight lines \cite{sakurai}. We now specify $\hat t$ explicitly for the rotationally free and for the rotationally diabatic transit through the grating.

\subsubsection*{Free Rotor Transit}

In a rotationally free transit through the grating the molecule rotates rapidly during the passage through the grating, $\tau_{\rm rot} \ll \tau_{\rm CM}$, and the rotational energy clearly exceeds the average potential. Then the transverse CM wavefunctions $\underline{\chi_{\ell}}$ are nearly constant during the transit and one can neglect the non-resonant terms in the Schr\"{o}dinger equation \eqref{eq:preeffschr} (rotating wave approximation),
\be \label{eq:effschr}
i \hbar \partial_t {\underline \chi}_\ell(x,t) = \left [ \hat H_{\rm CM}+ {\underline{ \underline V}}_{\ell \ell}\left ( x, v_z t \right ) \right ] {\underline \chi}_{\ell}(x,t).
\ee
The interaction potential is effectively diagonal in the angular momentum quantum numbers $\ell$ due to the fast molecular rotations. The corresponding expansion coefficients ${\underline \chi}_\ell(x,t)$ for different energies $\eps_\ell$ are mutually independent; however, in general the entries within each ${\underline \chi}_\ell(x,t)$ are coupled via Eq. \eqref{eq:effschr}.

In the eikonal approximation \cite{sakurai,kdtlinjp,stefanbeyond} the scattered state ${\underline \chi}'_{\ell}(x)$ behind the grating is according to the Schr\"{o}dinger equation \eqref{eq:effschr} related to the impinging state ${\underline \chi}_\ell(x)$ by ${\underline \chi}'_{\ell}(x) = {\underline {\underline T}}_{\ell}(x){\underline \chi}_{\ell}(x)$ where the grating transformation matrix is given by
\be \label{eq:transmat}
{\underline {\underline T}}_\ell(x) = \exp \left [ - \frac{i}{\hbar v_z} \int_{-\infty}^{\infty} \mathrm{d}{z}~ {\underline {\underline  V}}_{\ell \ell}(x,z) \right ].
\ee
It is a square matrix of dimension $\nu_\ell$. The corresponding grating transformation operator $\hat t$ can be expressed in terms of the matrix elements $t_\ell^{mm'} = ({\underline {\underline T}}_\ell)_{mm'}$ of the grating transformation matrix \eqref{eq:transmat} as
\be \label{eq:grattrafo}
\hat t = \sum_\ell \sum_{mm'} \left ( t_\ell^{mm'}(\hat x) \otimes \ketbra{\ell m}{\ell m'}  \right ) \vert t(\hat x, \hat \Omega) \vert .
\ee
Here, we included the aperture function $\vert t(x,\Omega) \vert \in \{0,1 \}$ describing the grating structure. Pure phase-gratings are characterized by $\vert t(x, \Omega) \vert = 1$ while for ideal interaction-free gratings the eikonal phase vanishes \cite{stefanbeyond}.

The grating transformation \eqref{eq:grattrafo} is valid for any interaction potential as long as the rotating wave approximation is justified. This is the case in most matter-wave experiments with large particles, such as near-field interference experiments with laser gratings \cite{kdtlinature,kdtlinjp,otimanjp} or far-field experiments with thick gratings \cite{colloquium}. In Sec. \ref{sec:example} we will specify the grating transformation \eqref{eq:grattrafo} for laser gratings. In addition, we note that in many practical cases the initial rotational state is a thermal mixture of angular momentum states and, hence, the grating transformation \eqref{eq:grattrafo} can be regarded as the thermal average of angular momentum dependent grating transformations $t_\ell^{mm'}(\hat x)$.

\subsubsection*{Diabatic Transit}

In a rotationally diabatic grating passage the interaction time $\tau_{\rm CM}$ is much smaller than the rotational period $\tau_{\rm rot}$ and, in a classical picture, the orientation $\Omega$ remains constant during the transit. In this case the quantum mechanical grating transformation gets diagonal in the orientational coordinates $\Omega$. In order to see this we note that for short transit times the rotating phases $\Delta_{\ell \ell'} t / \hbar \simeq 2 \pi t / \tau_{\rm rot}$ in the Schr\"{o}dinger equation \eqref{eq:preeffschr} can be neglected. This yields the coupled equations
\be \label{eq:effschr2}
i \hbar \partial_t {\underline \chi}_\ell(x,t) = \hat H_{\rm CM} {\underline \chi}_\ell(x,t) + \sum_{\ell'}{\underline {\underline V}}_{\ell \ell'}\left ( x, v_z t \right ) {\underline \chi}_{\ell'}(x,t),
\ee
which depend on the orientational DOFs only in parametric fashion, since these equations are independent of the rotational energy levels $\eps_\ell$. Defining the wavefunction,
\be
\Phi(x,\Omega,t) = \sum_{\ell} {\underline \chi}_\ell(x,t) \cdot {\underline \phi}_\ell(\Omega),
\ee
allows us to rewrite the coupled Eqs. \eqref{eq:effschr2} in the form
\be
i \hbar \partial_t \Phi(x, \Omega,t) = \left [ \hat H_{\rm CM} + V \left(x, v_z t,\Omega \right ) \right ] \Phi(x,\Omega,t),
\ee
which can now be solved in the eikonal approximation. The diabatic grating transformation operator $\hat t$, mapping the initial state $\Phi(x,\Omega)$ onto the outgoing state $\Phi'(x,\Omega)$, can now be written in the eikonal approximation as
\be \label{eq:grattrafo2}
\hat t = \vert t(\hat x, \hat \Omega) \vert \exp \left [ - \frac{i}{\hbar v_z} \int_{-\infty}^\infty \mathrm{d}{z}~ V(\hat x, z, \hat \Omega) \right ].
\ee
As anticipated, the diabatic grating transformation \eqref{eq:grattrafo2} is diagonal in the orientational DOFs $\Omega$. This coincides with the classical perception that the orientation is conserved during the diabatic passage through the grating, and is in contrast to the free rotor case \eqref{eq:grattrafo}, which is diagonal in the angular momentum quantum numbers $\ell$. The diabatic transformation \eqref{eq:grattrafo2} can be appropriate to describe the transit through ultra-thin material gratings, e.g. made of graphene.

\subsection{Classical Grating Transformation}

In order to identify genuine quantum effects in near-field matter-wave interferometry, it is necessary to compare the quantum interference signal to the classically expected shadow pattern of the grating \cite{kdtlinjp}. This moir\'{e}-type signal can in principle be obtained by solving Hamilton's equations of motion for a rotating molecule in the grating potential $V(x,v_z t,\Omega)$. However, the problem is significantly simplified by the classical analogue of the eikonal approximation.

The classical state of the rotating molecule approaching the grating is described by its phase space distribution function $f(x,p_x,\Omega,p_\Omega)$, where $p_\Omega$ denotes the vector of conjugate momenta to the angles $\Omega$. We seek the classical grating transformation, that maps the incoming state $f(x,p_x,\Omega,p_\Omega)$ onto the outgoing state $f'(x,p_x,\Omega,p_\Omega)$ \cite{kdtlinjp,klausdecoherence1}, i.e. the classical analogue of the grating transformation operator $\hat t$ mapping $\rho$ onto $\rho'$. We discuss the free rotor scenario first.

\subsubsection*{Free Rotor Transit}

For a rapidly rotating molecule, the CM motion is determined by the grating potential averaged over a rotational period \cite{goldstein}. The resulting eikonal momentum kick $\Delta p_x$ experienced by the molecule while passing through the grating reads as
\bea \label{eq:deltap}
\Delta p_x(x,\Omega,p_\Omega) & = & - \frac{1}{\tau_{\rm rot} v_z} \int_0^{\tau_{\rm rot}} \mathrm{d}{t'} ~\int_{- \infty}^\infty \mathrm{d}{z} \notag \\
 && \times \partial_x V \left [ x,z, \Omega(t') \right].
\eea
The transferred momentum \eqref{eq:deltap} is a function of both the transverse CM coordinate $x$ and of the initial orientation state $(\Omega,p_\Omega)$, that determines the rotational dynamics. In addition, it is reasonable to neglect the influence of the grating on the rotational dynamics, since the rotational energy is much higher than the average interaction potential. The free rotor approximation is in most practical cases well justified due to the high rotational temperature of the molecules in the experiments \cite{colloquium}.

With the help of the eikonal momentum kick \eqref{eq:deltap}, we can express the outgoing distribution $f'(x, p_x,\Omega,p_\Omega)$ as a CM momentum convolution of the impinging distribution $f(x, p_x, \Omega, p_\Omega)$,
\bea \label{eq:psconv}
f'(x,p_x,\Omega,p_\Omega) & = & \int \mathrm{d}{p_x'}~ T_{\rm cl}(x,p_x - p_x',\Omega, p_\Omega), \notag \\
 && \times f(x,p_x',\Omega, p_\Omega),
\eea
with the grating transformation function
\bea
T_{\rm cl} \left ( x, p_x, \Omega,p_\Omega \right ) = \vert t(x, \Omega) \vert \delta \left [p_x - \Delta p_x (x, \Omega, p_\Omega ) \right ].
\eea
The function $T_{\rm cl}(x,p_x,\Omega,p_\Omega)$ is the classical analogue of the quantum mechanical grating transformation operator $\hat t$. We remark that the transformation \eqref{eq:psconv} conserves the angular momenta $p_\Omega$, as it was the case in the quantum mechanical case \eqref{eq:transmat}. We will now specify the grating transformation function for the rotationally diabatic case.

\subsubsection*{Diabatic Transit}

In the case of a rotationally diabatic transit, the rotational period is much longer than the transit time and thus the orientation $\Omega$ of the molecule can be considered as being constant during the passage through the grating. The resulting CM momentum kick \eqref{eq:deltap} is
\be \label{eq:deltap2}
\Delta p_x(x,\Omega) = - \frac{1}{v_z} \int_{- \infty}^\infty \mathrm{d}{z}~ \partial_x V \left ( x,z, \Omega \right),
\ee
which is a function of the transverse CM position $x$ and of the orientation $\Omega$. In a similar fashion, we obtain the eikonal angular momentum kick $\Delta p_\Omega$,
\be
\Delta p_\Omega =  - \frac{1}{v_z} \int_{- \infty}^\infty \mathrm{d}{z}~ \partial_\Omega V \left ( x,z, \Omega \right),
\ee
which is also a function of $x$ and $\Omega$. The resulting grating transformation is now given by a CM momentum and an angular momentum convolution analogous to Eq. \eqref{eq:psconv} with the grating transformation function $T_{\rm cl}(x,p_x,\Omega,p_\Omega)$,
\bea \label{eq:tclas2}
T_{\rm cl} \left ( x, p_x, \Omega, p_\Omega \right ) & = & \vert t(x, \Omega) \vert \delta \left [p_x - \Delta p_x(x, \Omega ) \right ] \notag \\
 && \times \delta \left [p_\Omega - \Delta p_\Omega(x, \Omega ) \right ].
\eea
Having derived the quantum and classical grating transformations, we can next apply them to the molecular diffraction from standing wave laser gratings.

\section{Interference of Symmetric Top Molecules in the KDTLI} \label{sec:example}

Here, we first discuss the diffraction of rapidly rotating symmetric top molecules from laser gratings in order to illustrate the previously derived grating transformation. The obtained transformation operator is then used to determine the quantum fringe visibility as well as the classical shadow pattern in the Kapitza-Dirac-Talbot-Lau interferometer (KDTLI) \cite{kdtlinature,kdtlinjp,colloquium}. The KDTLI is a near-field interferometer consisting of three gratings that all share the same grating period $d$. The first and third grating are material masks, while the central one is a standing light wave. The transverse coherence of the incoming particle beam is prepared by the first material grating at distance $L$ in front of the standing wave. Diffraction occurs at the central grating and the signal is detected with the help of the third one at distance $L$ further downstream. The KDTLI is currently the working-horse for high mass interference experiments in Vienna \cite{colloquium}.

For the theoretical description of rotating molecules in the KDTLI it is a reasonable approximation to neglect the influence of the first and the third grating on the orientational DOFs because the diffraction relevant for interference takes place only at the central grating. The total transverse state operator $\rho$ behind the third grating can be obtained by successively applying the grating transformations of the three gratings as well as the intermediate unitary evolutions \cite{stefanbeyond,kdtlinjp}. Finally, the orientational DOFs are traced out in order to obtain the interference pattern on the screen.

\subsection{Standing Wave Grating Transformation}

We now evaluate the grating transformation $\hat t$ for symmetric top molecules traversing a standing-wave laser grating. This operator will then be used to calculate the near-field interference pattern of symmetric molecules in the KDTLI, but it can also be applied to other situations. We consider a polarizable, symmetric molecule with moments of inertia $I = I_1 = I_2$ and $I_3$ ($I / I_3 \geq 1$ for prolate particles and $1/2 < I / I_3 < 1$ for oblate discs), that is diffracted from a Gaussian standing laser wave of wavelength $\lambda$. The laser wave is linearly polarized in the direction ${\bf n}$ and acts as a pure phase grating, $\vert t(x,\Omega) \vert = 1$. The  intensity of the Gaussian standing laser beam averaged over one optical cycle is given by \cite{kdtlinjp}
\be \label{eq:int}
I(x,z) = \frac{8 P}{\pi w_y w_z} \exp \left ( - \frac{2 z^2}{w_z^2} \right ) \sin^2 \left ( \pi \frac{x}{d} \right ),
\ee
where $d = \lambda / 2$ is the grating period, $P$ the laser power and $w_z$ is the beam waist in $z$-direction. Since the extension of the incoming particle beam in $y$-direction is usually small compared to the beam waist in $y$-direction \cite{kdtlinature} it is natural to neglect the $y$-dependence of the intensity \eqref{eq:int}.

Denoting by $\alpha_\|$ and $\alpha_\bot$ the two independent components of the polarizability tensor of the particle (along its symmetry axis and perpendicular to it, respectively), the grating potential can be expressed as \cite{kdtlinjp,averbukh2010,averbukh2011a,averbukh2011b} \footnote{We assume here that the extension of the molecule is much less than the period of the standing wave; otherwise the use of a polarizability tensor would be invalid.}
\bea \label{eq:pot1}
V \left ( x, z, \theta \right ) & = & -\frac{4 P}{\pi \eps_0 c w_z w_y} \exp \left ( - \frac{2 z^2}{w_z^2} \right ) \notag \\
 && \times \left ( \alpha_\| -  \Delta \alpha \sin^2 \theta \right )\sin^2 \left ( \pi \frac{x}{d} \right ) .
\eea
Here $\Delta \alpha = \alpha_\| - \alpha_\bot$ is the polarizability anisotropy of the molecule ($\Delta \alpha > 0$ for prolate particles) and $\theta$ is the nutation angle with respect to the field polarization ${\bf n}$. The grating transformation can be safely determined in the free rotor approximation, Eq. \eqref{eq:grattrafo}, since the laser beam waist along the flight direction is approximately $w_z = 20$ $\mu$m \cite{kdtlinature,kdtlinjp}. For an exemplary molecule (diazobenzene with mass $M \simeq 1030$ amu and length $L_{\rm mol} \simeq 3.5$ nm) with velocity $v_z = 100$ m s$^{-1}$ and rotational temperature $T = 600$ K, we thus have $\tau_{\rm rot} / \tau_{\rm CM} \sim 10^{-4}$.

In order to calculate the grating transformation matrix \eqref{eq:transmat} we need to specify the grating potential matrices, whose particular form depends on the symmetry of the rotor. Here we restrict our discussion to symmetric top molecules for reasons of simplicity. The eigenstates $\ket{\ell m k}$ of the symmetric rotor (with classical Hamilton function \eqref{eq:hamrot}, see appendix) are labeled by the three quantum numbers $\ell \in \NN_0$, $m \in \{-\ell,\ldots, \ell \}$ and $k \in \{ -\ell, \dots, \ell \}$ with eigenenergies $\eps_{\ell k}$ \cite{edmonds,brink}. Denoting by $\varphi \in [0,2 \pi)$, $\theta \in [0,\pi)$ and $\psi \in [0,2 \pi)$ the three Euler angles with respect to the field polarization ${\bf n}$ ($z$-$y'$-$z''$ convention), the configuration space representation of the states $\ket{\ell m k}$ can be given explicitly in terms of the Wigner $D$-matrices. In particular, the eigenfunctions are $\braket{\Omega}{\ell m k} = \sqrt{2 \ell +1}/ (2 \pi \sqrt{2} ) D^{\ell *}_{m k}(\Omega)
$ \cite{edmonds,brink}, where $D^\ell_{mk}(\Omega)$ is related to the (small) Wigner $d$-matrix by $D_{mk}^\ell(\Omega) = e^{-im\varphi} d_{mk}^\ell(\theta) e^{-i k \psi}$. We remark that the (small) Wigner $d$-matrices $d_{mk}^\ell(\theta)$ are real due to the employed definition of the Euler angles \cite{sakurai}. The basis kets $\ket{\ell m k}$ are complete and orthonormal with respect to the infinitesimal volume element $\mathrm{d}{\Omega} = \mathrm{d}{\varphi}~ \mathrm{d}{\theta}~ \mathrm{d}{\psi}~ \sin \theta$.

The interaction potential \eqref{eq:pot1} is a function of the nutation angle $\theta$ only and thus the quantum numbers $m$ and $k$ are conserved. The resulting grating potential is diagonal in all three quantum numbers $\ell$, $m$, $k$, and the diagonal elements can be given with the help of the expectation values $Q_{\ell m k} := \matel{\ell m k}{\sin^2 \hat \theta}{\ell m k}$.  Expressing these expectation values in terms of Wigner $D$-matrices \cite{seideman05} and using the properties of the Wigner $3$-j symbol \cite{edmonds,brink} yields the algebraic form
\bea \label{eq:qlmk}
Q_{\ell m k} & = & \frac{1}{2} + \frac{1}{2} \frac{(2m)^2 + (2 k )^2 - 1}{(2 \ell -1 )(2 \ell + 3)} \notag \\
 && - \frac{3}{2} \frac{(2 m k)^2}{\ell ( \ell + 1)  ( 2 \ell -1 )(2 \ell +3)}.
\eea
We remark that in the limit of a linear rotor, $I / I_3 \to \infty$ and thus $k = 0$, the well-known expectation value \cite{averbukh2010}
\be\label{eq:qlm}
Q_{\ell m 0} = \frac{1}{2}  + \frac{1}{2} \frac{(2 m)^2 - 1}{(2 \ell - 1)(2 \ell + 3)},
\ee
of the linear rigid rotor is recovered.

We are now ready to identify the grating transformation operator for symmetric top molecules traversing a standing wave laser grating by inserting the expectation value \eqref{eq:qlmk} into the grating transformation matrix \eqref{eq:transmat}. The resulting operator \eqref{eq:grattrafo} is
\bea \label{eq:transmatlaser}
\hat t & = & \sum_{\ell = 0}^\infty \sum_{m, k = - \ell}^\ell \exp \left [ i \phi_0 \left ( 1 - \frac{\Delta \alpha}{\alpha_\|} Q_{\ell mk} \right ) \sin^2 \left ( \pi \frac{\hat x}{d} \right ) \right ] \notag \\
 && \otimes \ketbra{\ell m k}{\ell m k},
\eea
where $\phi_0 = 4 \alpha_\| P / \eps_0 c\hbar w_y v_z \sqrt{2 \pi}$ is the eikonal phase \cite{kdtlinjp} defined with the polarizability $\alpha_\|$ along the symmetry axis of the molecule. It is important to note that the eikonal phase imprinted on a particle during the grating passage depends on all of its angular momentum quantum numbers $\ell$, $m$ and $k$. The final signal, obtained by a trace over the orientational DOFs, is thus an average over signals from different grating transformations \eqref{eq:transmatlaser}, each weighted with the probability of the corresponding angular momentum state $\ket{\ell m k}$. This matches with the fact that the classical deflection angle of molecules traversing an electrostatic field depends on the angular momentum of the deflected particle \cite{averbukh2010,averbukh2011a,averbukh2011b}. For nearly isotropic particles, $\Delta \alpha / \alpha_\| \ll 1$, the transformations \eqref{eq:transmatlaser} are all equal and the average over angular 
momentum states can be neglected.

\subsection{Quantum and Classical Fringe Visibility}

Having specified the grating transformation \eqref{eq:transmatlaser}, the quantum fringe visibility $\mathcal V$ of symmetric top molecules in the KDTLI can be calculated by applying the transformation \eqref{eq:grattrafo} with \eqref{eq:transmatlaser} to the quantum phase space formalism presented in \cite{kdtlinjp}. The common period of all three gratings is denoted by $d$ and the de Broglie wavelength of the incoming rod by $\lambda_{\rm dB} = h / M v_z$. The Talbot length, the characteristic length scale in near-field interferometry \cite{colloquium}, is $L_{\rm T} = d^2 / \lambda_{\rm dB}$. A straightforward calculation yields the sinusoidal quantum fringe visibility
\bea \label{eq:vis}
{\mathcal V} & = & 2 \, {\rm sinc}^2 ( \pi f)  \sum_{\ell = 0}^\infty \sum_{m,k = - \ell}^\ell p_{\ell m k}  \notag \\
 && \times J_2 \left [\phi_0 \left ( 1 - \frac{\Delta \alpha}{\alpha_\|} Q_{\ell m k} \right ) \sin \left ( \pi \frac{L}{L_{\rm T}} \right ) \right ],
\eea
where $f$ is the opening fraction of the first and the third grating, $J_2( \cdot )$ is the second order Bessel function of the first kind, $L$ is the distance between the gratings and $p_{\ell m k}$ is the statistical weight of the angular momentum state $\ket{\ell m k}$. The interference contrast can be regarded as the average of point particle visibilities with $(\ell,m,k)$-dependent eikonal phases and weights $p_{\ell m k}$.

\begin{figure}
 \centering
 \includegraphics[width = 80mm]{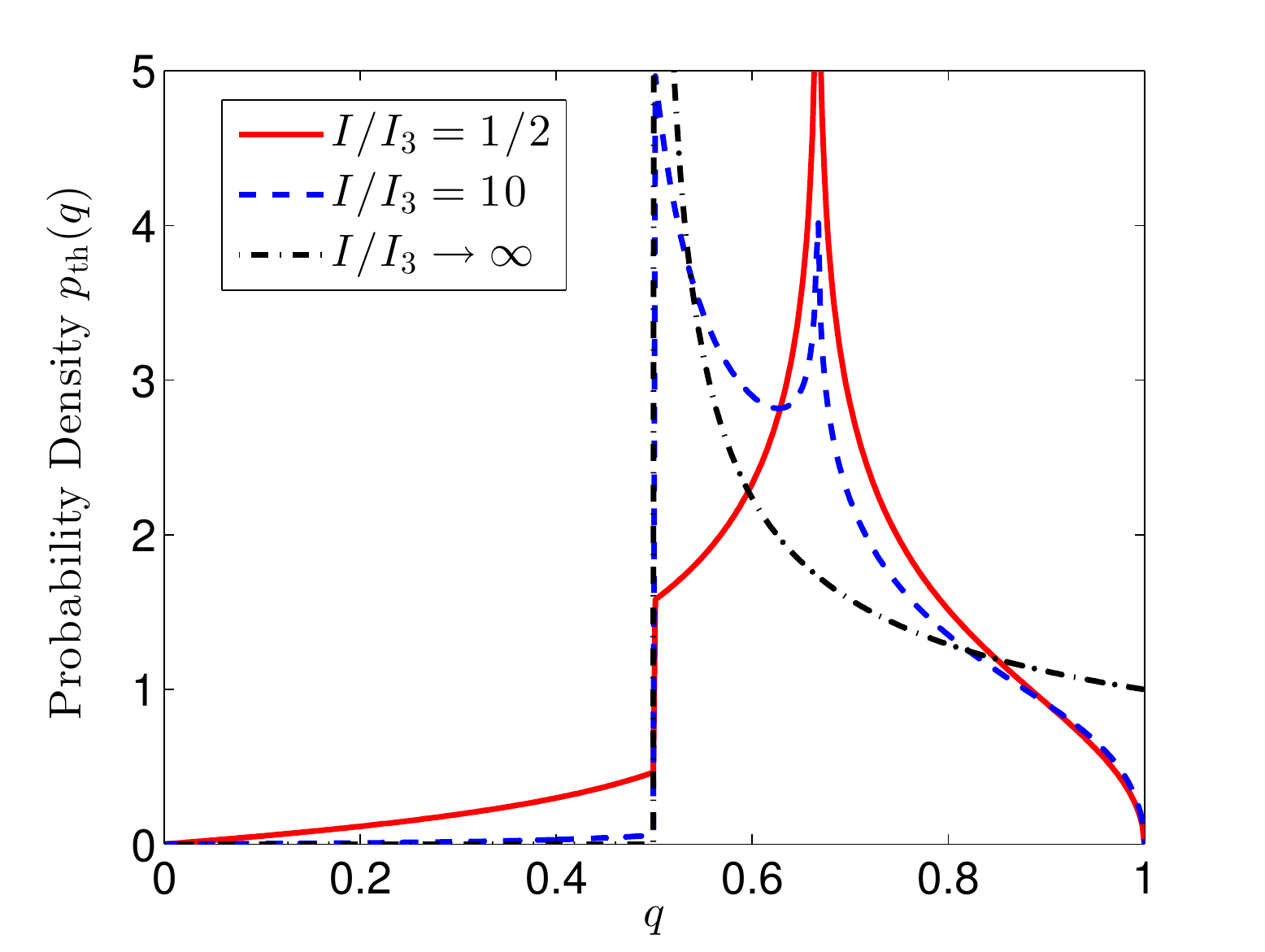}
 \caption{(Color online) The thermal distribution $p_{\rm th}(q)$, Eq. \eqref{eq:pth}, of the temporal average $q$ of $\sin^2 \theta(t)$ for the free symmetric rotor with different moments of inertia, $I / I_3 = 1/2$ (solid line), $I / I_3 = 10$ (dashed line) and $I / I_3 \to \infty$ (dot-dashed line).}\label{fig:rotdist}
\end{figure}

Since the molecules are emitted from a thermal source into vacuum the rotational DOFs follow a thermal distribution, $p_{\ell m k} \sim \exp ( - \eps_{\ell k} / k_{\rm B} T )$, at a very high temperature, $k_{\rm B} T \gg \hbar^2 / I$. Then the sum over angular momenta in Eq. \eqref{eq:vis} can be replaced by the integral over the corresponding classical distribution \cite{brink} and Eq. \eqref{eq:vis} is further simplified. In particular, we denote by $p_{\rm th}(q)$ the probability density of the variable $q = Q(E_{\rm rot}, p_\varphi, p_\psi)$, where $Q(E_{\rm rot},p_\varphi,p_\psi)$ is the classical free temporal mean value of $\sin^2 \theta(t)$ depending on the conserved rotational energy $E_{\rm rot}$ and on the canonical momenta $p_\varphi$ and $p_\psi$ of $\varphi$ and $\psi$ rotations, respectively. A simple expression for $Q(E_{\rm rot},p_\varphi,p_\psi)$, as well as for the thermal distribution $p_{\rm th}(q)$, is derived in the appendix. This probability density is depicted in Fig.~\ref{fig:rotdist} and reads
\bea \label{eq:pth}
p_{\rm th}(q) & = & \sqrt{\frac{I}{3 I_3}} \int_{\zeta(q)} \mathrm{d}{u}~ \left [1 - \left (1 - \frac{I}{I_3} \right) u^2 \right ]^{- 3 / 2} \notag \\
 && \times \left [\left (u^2 - \frac{1}{3} \right ) \left ( u ^2 + 1 - 2q \right ) \right ]^{-1/2}.
\eea
where the integral must be taken over the union of two intervals, $\zeta(q) = \zeta_1(q) \cup \zeta_2(q)$. The first interval $\zeta_1(q) = [0,\sqrt{\mathrm{min}[A(q)]}]$, where $A(q) = \{ 1/3, 2 q - 1, 1 - q\}$. This contribution to the distribution \eqref{eq:pth} vanishes for $q \leq 1/2$. The second interval $\zeta_2(q) = [\sqrt{\mathrm{max}[A(q)]},1]$. The distribution \eqref{eq:pth} depends on the fraction $I / I_3$ only and is independent of the thermal energy $k_{\rm B} T$.

In Fig.~\ref{fig:rotdist} we show the distribution \eqref{eq:pth} of the symmetric rotor for the oblate limit ($I / I_3 = 1/2$), a prolate particle ($I / I_3 = 10$) and the linear rotor ($I / I_3 \to \infty$). Similar figures were obtained numerically in \cite{averbukh2011a}. For finite $I / I_3$ the probability density \eqref{eq:pth} is discontinuous at $q = 1/2$, which follows from the definition of the set $\zeta(q)$ and it diverges at $q = 2/3$, as can be observed directly from Eq. \eqref{eq:pth}. In the limit of the linear rotor, $I / I_3 \to \infty$, the established  \cite{averbukh2010} form  $p_{\rm th}(q) = 1 / \sqrt{2 q - 1}$ is recovered.

Using the probability density \eqref{eq:pth} the quantum fringe visibility $\mathcal V$ takes on its final form
\bea \label{eq:vis2}
{\mathcal V} & = & 2 \, {\rm sinc}^2 ( \pi f) \int_0^1 \mathrm{d}{q}~ p_{\rm th}(q) \notag \\
 && \times J_2 \left [\phi_0 \left ( 1 - \frac{\Delta \alpha}{\alpha_\|} q \right ) \sin \left ( \pi \frac{L}{L_{\rm T}} \right ) \right ] .
\eea
In order to identify genuine quantum interference effects we must compare the visibility \eqref{eq:vis2} to the visibility $\mathcal{V}_{\rm cl}$ of the classical shadow pattern \cite{kdtlinjp}, which is most conveniently calculated with the help of the phase space grating transformations \eqref{eq:psconv}. The classical momentum kick \eqref{eq:deltap} transferred to the molecule by the grating potential \eqref{eq:pot1} is
\bea
\Delta p_x(x,E_{\rm rot}, p_\varphi)  &=& \frac{\pi \hbar \phi_0}{d}\left [ 1 - \frac{\Delta \alpha}{\alpha_\|} Q(E_{\rm rot},p_\varphi,p_\psi) \right ] \notag \\
 && \times \sin \left ( 2 \pi \frac{x}{d} \right ).
\eea
Following the treatment in \cite{kdtlinjp} one obtains the classical fringe visibility
\bea \label{eq:nucl}
\mathcal{V}_{\rm cl} & = & 2 \, \mathrm{sinc}^2 ( \pi f) \int_{0}^1 \mathrm{d}{q}~ p_{\rm th}(q) \notag \\
 & & \times J_2 \left [\left ( 1 - \frac{\Delta \alpha}{\alpha_\|} q \right ) \frac{\phi_0 \pi L}{L_{\rm T}} \right ],
\eea
where we assumed the orientational DOFs to be thermally distributed.

\begin{figure}
\centering
 \includegraphics[width = 80mm]{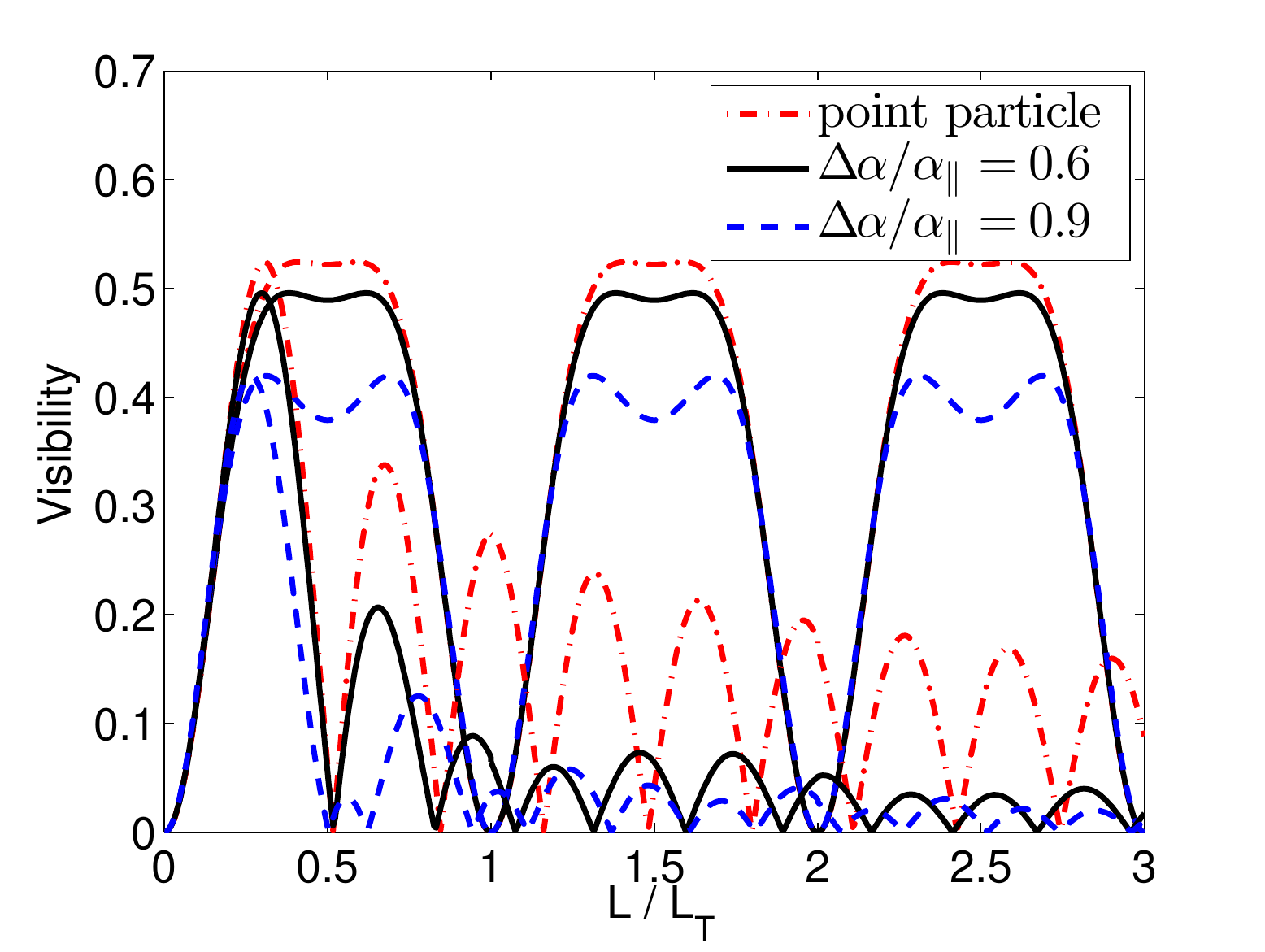}
 \caption{(Color online) Quantum (upper curves) and classical (lower curves) absolute sinusoidal fringe visibility as a function of relative separation $L / L_T$ for a linear molecule in the KDTLI for three different values of $\Delta \alpha / \alpha_\|$.}\label{fig:vis3}
\end{figure}

\begin{figure}
 \centering
 \includegraphics[width = 80mm]{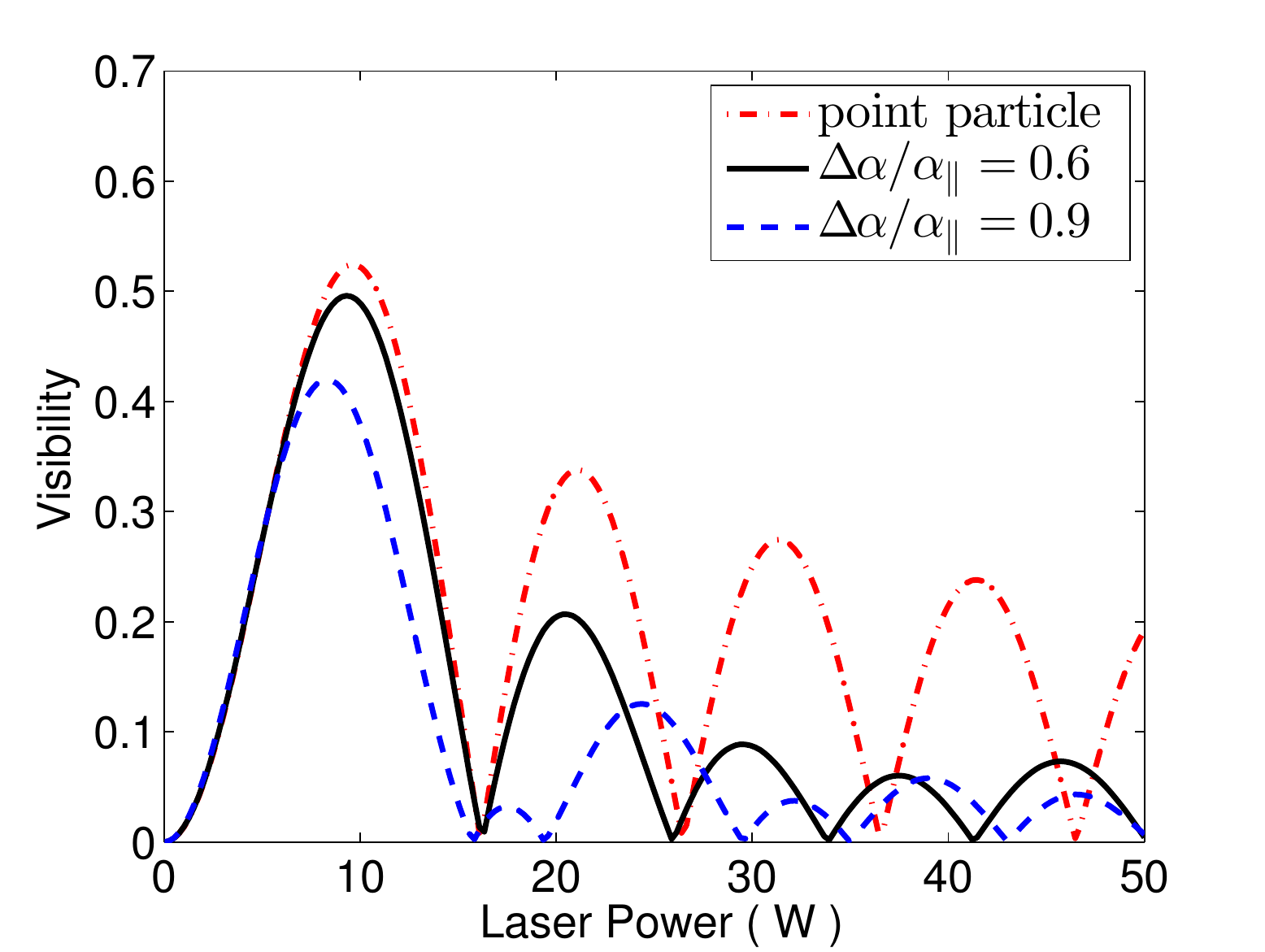}
 \caption{(Color online) Absolute value of the sinusoidal quantum fringe visibility \eqref{eq:vis} as a function of laser power for a linear rigid molecule in the KDTLI for three different values of $\Delta \alpha / \alpha_\|$.}\label{fig:vis1}
\end{figure}

The classical visibility \eqref{eq:nucl} decays as $\sqrt{L_{\rm T} / L}$ with increasing grating separation $L$ (and decreasing particle velocity $v_z$) while the quantum visibility \eqref{eq:vis2} is periodic in $L / L_{\rm T}$ \cite{kdtlinature,kdtlinjp}. This can be used to discriminate between genuine quantum behavior and classical shadow effects \cite{kdtlinature} as illustrated in Fig.~\ref{fig:vis3} for the linear rotor $I / I_3 \to \infty$. In Figs.~\ref{fig:vis3} and \ref{fig:vis1} we consider an exemplary molecule ($M  = 1030$ amu, $L_{\rm mol} = 3.5$ nm, $\overline \alpha = 4 \pi \eps_0 \times 50$ \AA$^3$, $v_z = 100$ m s$^{-1}$, $I / I_3 \simeq \infty$ \cite{kdtlinature}) traversing the KDTLI ($d = 266$ nm, $L / L_T = 0.5$, $f = 0.42$, $w_z = 20$ $\mu$m \cite{kdtlinature,kdtlinjp}) for three different relative anisotropies $\Delta \alpha / \alpha_\|$.

An experimentally observable signature of the orientational DOFs can be found in the absolute value of the quantum fringe visibility \eqref{eq:vis2} as a function of laser power $P$. This is illustrated in Fig.~\ref{fig:vis1}. While the value of subsequent maxima in the visibility are strictly decreasing for spherical particles, this is not the case for non-spherical molecules. For large relative anisotropies $\Delta \alpha / \alpha_\|$ the thermal average over angular momentum states leads to the appearance of additional side peak between the major recurrences. This is a signature of the orientational DOFs of the molecule traversing the grating.

\begin{figure}
 \centering
 \includegraphics[width = 80mm]{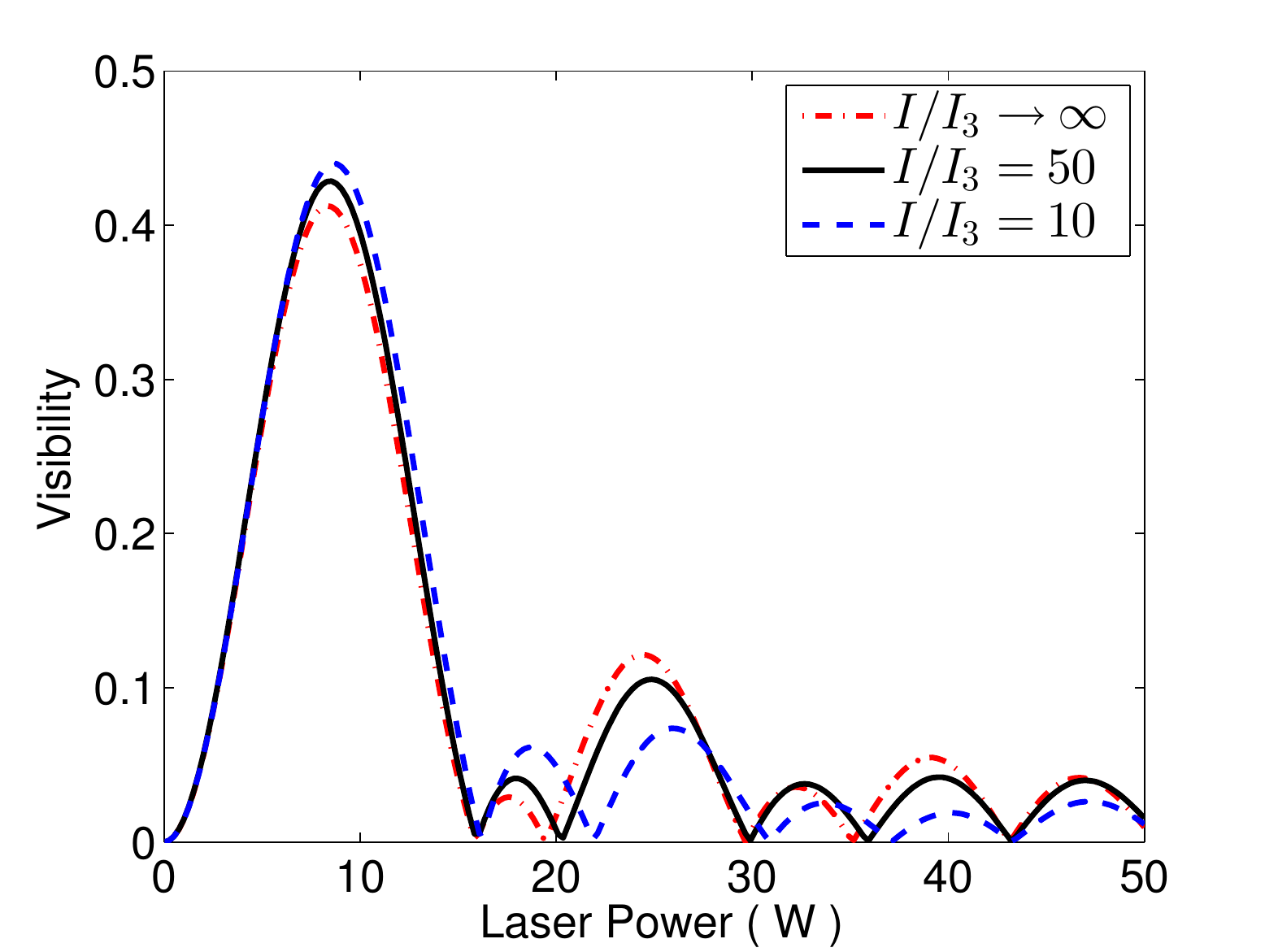}
 \caption{(Color online) Absolute value of the sinusoidal quantum fringe visibility \eqref{eq:vis2} as a function of laser power for the prolate symmetric top in the KDTLI for different values of $I / I_3$ ($\Delta \alpha / \alpha_\| = 0.9$).}\label{fig:vis5}
\end{figure}

In Fig.~\ref{fig:vis5} we show the absolute value of the quantum fringe visibility \eqref{eq:vis2} as a function of laser power for three differently shaped prolate molecules. All other parameters are as for the linear molecules of Fig.~\ref{fig:vis1} ($\Delta \alpha / \alpha_\| = 0.9$). While the visibility coincides with the visibility of the linear molecule for large ratios $I / I_3$, it approaches the behavior of spherically symmetric object for $I / I_3 \to 1$. The signature of the orientational DOFs discussed above is most pronounced for linear molecules but can be observed for all prolately shaped molecules.

\section{Conclusion} \label{sec:conc}

We extended the theory of matter-wave interferometry to large, non-spherical particles by accounting for the influence of the rotational dynamics. In particular, we derived the grating transformation operator for the rotationally free and for the rotationally diabatic transit. This operator describes the modification of the transverse quantum state due to the orientation-dependent interaction with the grating. In addition, the classical shadow pattern was derived in order to provide the tools required for the identification of genuine quantum effects in near-field matter-wave interferometry.

If the molecule rotates rapidly with high energy, the transit is rotationally free and the grating transformation depends only on the angular momentum of the impinging particle. On the other hand, if the transit time is much shorter than the average rotational period, the grating transformation depends on the orientation of the particle.

We worked out the grating transformation for symmetric top molecules traversing a standing-wave laser grating, and we showed how it enters the description of symmetric top particles in the Kapitza-Dirac-Talbot-Lau interferometer as performed at the University of Vienna \cite{kdtlinature}. In these experiments the typical transit time exceeds the rotational period by orders of magnitude, making the transit rotationally free. A signature of the rotational dynamics was pointed out in the predicted quantum fringe visibility as a function of laser power. We also derived a closed-form expression for the distribution of deflection angles in classical deflection experiments with symmetric top molecules, as required in this context.

\section{Acknowledgments}

We acknowledge support by the European Commission within NANOQUESTFIT (No. 304886).

\appendix

\section{Rotation Statistics of Symmetric Top Molecules} \label{app:dist}

In this appendix we present a derivation of the thermal probability distribution $f_{\rm th}(r)$ of classical realizations $r$ of the temporal mean value of $\cos^2 \theta(t)$ for symmetric top molecules, as required in Sec. \ref{sec:example}. This distribution is of central interest in the theory of molecular deflection experiments because it translates directly into the distribution of deflection angles \cite{averbukh2010,seideman05,stapelfeldt03,denschlag14} and it has been evaluated numerically with the help of Monte-Carlo methods in \cite{averbukh2011a}. The distribution \eqref{eq:pth} of realizations $q$ of $\sin^2 \theta(t)$ can be trivially obtained by substituting $r =  1 - q$ in $f_{\rm th}(r)$.

We follow the notation of Sec. \ref{sec:example} and denote the moments of inertia of the symmetric top by $I_1 = I_2 \equiv I$ and $I_3$. The three orientational DOF $\Omega = (\varphi,\theta, \psi)$ are the Euler angles in the $z$-$y'$-$z''$ convention, and the respective conjugate momenta are $p_\Omega = (p_\varphi,p_\theta,p_\psi)$. The classical Hamilton function $H_{\rm rot}(\Omega, p_\Omega)$ of the free symmetric top is given by \cite{goldstein}
\be \label{eq:hamrot}
H_{\rm rot}(\Omega,p_\Omega) = \frac{1}{2 I} \left ( \frac{(p_\varphi - p_\psi \cos \theta)^2}{\sin^2 \theta} + p_\theta^2 \right ) + \frac{p_\psi^2}{2 I_3},
\ee
and there are three conserved quantities: the total energy $E_{\rm rot} = H_{\rm rot}(\Omega,p_\Omega)$, as well as the angular momenta $p_\varphi$ and $p_\psi$. Hence, the equations of motion are integrable and the period of $\theta$-rotations acquires the form
\be \label{eq:period}
\tau_{\rm rot} = \frac{2 \pi I}{\sqrt{ 2 E_{\rm rot} I + p_\psi^2 \left ( 1 - I / I_3 \right )}}.
\ee
In the limit of very prolate molecules, $I / I_3 \to \infty$, $\psi$-rotations do not contribute, $p_\psi = 0$, and the period $\tau_{\rm rot}$ approaches the period of the linear rotor, $\tau_{\rm rot} \to \pi \sqrt{2 I / E_{\rm rot}}$.

Separating Hamilton's equations with \eqref{eq:hamrot} yields the temporal average of $\cos^2 \theta(t)$ over one rotational period \eqref{eq:period},
\bea \label{eq:average}
R(E_{\rm rot}, p_\varphi, p_\psi) & = & \frac{1}{\tau_{\rm rot}} \int_0^{\tau_{\rm rot}} \mathrm{d}{t'}~ \cos^2 \theta(t') \notag \\
 & = & \frac{1}{2} - \frac{1}{2} \frac{p_\varphi^2 + p_\psi^2}{2 E_{\rm rot} I + p_\psi^2 \left ( 1 - I / I_3 \right )} \notag \\
  && + \frac{3}{2} \left ( \frac{p_\varphi p_\psi}{2 E_{\rm rot} I + p_\psi^2 \left ( 1 - I / I_3 \right )} \right )^2.
\eea
We observe that the quantum mechanical expectation value $R_{\ell mk} = 1 - Q_{\ell mk}$, Eq. \eqref{eq:qlmk}, closely resembles this expression.

A more compact expression for the temporal average \eqref{eq:average} is obtained in terms of the relative frequencies of $\varphi$- and $\psi$-rotations
\be \label{eq:u12}
u_1  = \frac{p_\varphi\tau_{\rm rot} }{2 \pi I} \quad \text{and} \quad u_2 =  \frac{ p_\psi\tau_{\rm rot}}{2 \pi I},
\ee
which satisfy $-1 \leq u_{1},u_2 \leq 1$. In particular, introducing the angular momentum scale $p_{\rm rot} = 2 \pi I / \tau_{\rm rot}$ we observe that $\vert p_{\varphi,\psi} \vert \leq p_{\rm rot}$. Then, the average \eqref{eq:average} can be written as
\bea \label{eq:r2}
\widetilde R (u_1,u_2) = \frac{1}{2} - \frac{1}{2} (u^2_1 + u^2_2) + \frac{3}{2} u^2_1 u^2_2,
\eea
which will be of advantage in what follows.

If the orientational DOFs are distributed in phase space according to the thermal distribution
\bea \label{eq:thermdist}
p_{\rm th}(\Omega,p_\Omega) & = & \frac{1}{Z}\exp \left [ - H_{\rm rot}(\Omega,p_\Omega) /k_{\rm B} T \right ],
\eea
with the partition function $Z = 16 \pi^3 I  k_{\rm B} T\sqrt{2 \pi I_3 k_{\rm B} T}$ \cite{townes}, the distribution $f_{\rm th}(r)$ of $r = R(E_{\rm rot},p_\varphi,p_\psi)$ is given by
\be \label{eq:pq}
f_{\rm th}(r) = \int \mathrm{d}{\Gamma}~ \delta \left [ r - R(E_{\rm rot},p_\varphi,p_\psi) \right ] p_{\rm th}(\Omega,p_\Omega),
\ee
where the integral covers the whole phase space ($\mathrm{d}{\Gamma} = \mathrm{d}{\varphi}~ \mathrm{d}{\theta}~ \mathrm{d}{\psi}~ \mathrm{d}{p_\varphi}~ \mathrm{d}{p_\theta}~ \mathrm{d}{p_\psi}$). On the other hand, one can also write the distribution $f_{\rm th}(r)$, Eq. \eqref{eq:pq}, in terms of the probability density $q_{\rm th}(u_1,u_2)$,
\bea \label{eq:pu1u2}
q_{\rm th}(u_1,u_2) & = & \int \mathrm{d}{\Gamma}~ \delta \left ( u_1 - \frac{p_\varphi \tau_{\rm rot}}{2 \pi I} \right )\notag \\
 && \times \delta \left ( u_1 - \frac{p_\psi \tau_{\rm rot}}{2 \pi I} \right )  p_{\rm th}(\Omega,p_\Omega),
\eea
according to
\be \label{eq:pr2}
f_{\rm th}(r) = \int_{-1}^1 \mathrm{d}{u_1}~ \int_{-1}^1 \mathrm{d}{u_2}~ \delta \left [ r - \widetilde R(u_1,u_2) \right ] q_{\rm th}(u_1,u_2).
\ee

The integral \eqref{eq:pu1u2} can be evaluated by defining the dimensionless quantities $\omega_\chi = ( p_\varphi - p_\psi \cos \theta) / \sin \theta \sqrt{I k_{\rm B} T}$, $\omega_\theta = p_\theta / \sqrt{I k_{\rm B} T}$ and $\omega_\psi = p_\psi / \sqrt{I k_{\rm B} T}$ and transforming the vector $(\omega_\chi,\omega_\theta,\omega_\psi)$ to spherical coordinates, $(\omega, \xi, \eta)$ (where $\omega \equiv p_{\rm rot} / I$). This gives
\bea \label{eq:pu1u22}
q_{\rm th}(u_1,u_2) & = & \frac{1}{4} \sqrt{\frac{I}{I_3}}\left [1 - \left (1 - \frac{I}{I_3} \right) u_2^2 \right ]^{- 3 / 2},
\eea
which is independent of the temperature $T$, uniformly distributed in $u_1$, a function of the ratio $I / I_3$ only, and normalized by construction.

Finally, inserting the distribution \eqref{eq:pu1u22} into expression \eqref{eq:pr2} yields the thermal distribution $f_{\rm th}(r)$,
\bea \label{eq:prfinal}
f_{\rm th}(r) & = & \sqrt{\frac{I}{3I_3}} \int_{\zeta(r)} \mathrm{d}{u}~ \left [1 - \left (1 - \frac{I}{I_3} \right) u^2 \right ]^{- 3 / 2} \notag \\
 && \times \left [\left (u^2 - \frac{1}{3} \right ) \left ( u ^2 + 2r - 1 \right ) \right ]^{-1/2}.
\eea
Here, the set $\zeta(r) \subset [0,1]$ denotes the fraction of the unit interval $[0,1]$ where the integrand is real. In particular, for arbitrary $r \in [0,1]$, $\zeta(r) = \zeta_1(r) \cup \zeta_2(r)$ consists of two intervals $\zeta_{1,2}(r)$: the first is $\zeta_1(r) = [0,\sqrt{{\rm min}[ A(r)]}]$, where $A(r) = \{1/3, 1 - 2 r,r\}$. This contribution vanishes for $r \geq 1/2$. The second interval $\zeta_2(r)$ is $\zeta_2(r) = [\sqrt{\mathrm{max} [A(r)]},1]$ to one. The probability density function \eqref{eq:pth} of the temporal average $q = \langle \sin^2 \theta(t) \rangle$, as required in Sec. \ref{sec:example}, is obtained by the substitution $q = 1 - r$. Finally we note that the distribution $f_{\rm th}(r)$ \eqref{eq:prfinal} approaches in the rigid rotor limit $I / I_3 \to \infty$ (in agreement with \cite{averbukh2010})
\be \label{eq:classdist}
f_{\rm th}(r) = \frac{1}{\sqrt{1- 2r}}.
\ee

\bibliographystyle{unsrt}
\bibliographystyle{apsrev4-1}

\end{document}